\documentclass[aps,prb,twocolumn,superscriptaddress,floatfix]{revtex4-1}

\usepackage{graphicx}
\usepackage{dcolumn}
\usepackage{bm}
\usepackage[hidelinks]{hyperref}
\usepackage{siunitx}
\usepackage[version=4]{mhchem}
\usepackage{xcolor}
\usepackage{lipsum}

\newcommand{\mob}[1]{\SI[per-mode=symbol]{#1}{\square\centi\meter\per\volt\per\second}}

\DeclareSIUnit\sq{\ensuremath{\Box}}

\begin{document}

\preprint{xxxxxx}

\title{Characterising Quantum Devices at Scale with Custom Cryo-CMOS}

\author{S. J. Pauka}
    \thanks{These authors contributed equally to this work}
    \affiliation{ARC Centre of Excellence for Engineered Quantum Systems, School of Physics, The University of Sydney, Sydney, NSW 2006, Australia.}
\author{K. Das}
    \thanks{These authors contributed equally to this work}
    \affiliation{Microsoft Quantum Sydney, The University of Sydney, Sydney, NSW 2006, Australia.}
\author{J. M. Hornibrook}
    \affiliation{Microsoft Quantum Sydney, The University of Sydney, Sydney, NSW 2006, Australia.}
\author{G. C. Gardner}
    \affiliation{Birck Nanotechnology Center, Purdue University, West Lafayette, IN 47907, USA.}
    \affiliation{Microsoft Quantum Purdue, Purdue University, West Lafayette, IN 47907, USA.}
\author{M. J. Manfra}
    \affiliation{Department of Physics and Astronomy, Purdue University, West Lafayette, IN 47907, USA.}
    \affiliation{Birck Nanotechnology Center, Purdue University, West Lafayette, IN 47907, USA.}
    \affiliation{Microsoft Quantum Purdue, Purdue University, West Lafayette, IN 47907, USA.}
    \affiliation{School of Materials Engineering and School of Electrical and Computer Engineering, Purdue University, West Lafayette, IN 47907, USA.}
\author{M. C. Cassidy}
    \affiliation{Microsoft Quantum Sydney, The University of Sydney, Sydney, NSW 2006, Australia.}
\author{D. J. Reilly}
    \email{david.reilly@sydney.edu.au}
    \affiliation{ARC Centre of Excellence for Engineered Quantum Systems, School of Physics, The University of Sydney, Sydney, NSW 2006, Australia.}
    \affiliation{Microsoft Quantum Sydney, The University of Sydney, Sydney, NSW 2006, Australia.}

\date{\today}

\begin{abstract}
We make use of a custom-designed cryo-CMOS multiplexer (MUX) to enable multiple quantum devices to be characterized in a single cool-down of a dilution refrigerator. Combined with a packaging approach that integrates cryo-CMOS chips and a hot-swappable, parallel device test platform, we describe how this setup takes a standard wiring configuration as input and expands the capability for batch-characterization of quantum devices at milli-Kelvin temperatures and high magnetic fields. The architecture of the cryo-CMOS multiplexer is discussed and performance benchmarked using few-electron quantum dots and Hall mobility-mapping measurements.
\end{abstract}

\maketitle


\section{\label{sec:intro}Introduction}
Developing large-scale quantum machines brings new and distinct challenges not apparent in early-demonstration experiments with single devices or few-qubit systems \cite{2019arXiv190611146R,Veldhorst2017,PhysRevApplied.3.024010}. Although many fundamental scientific barriers stand in the path to scale-up, significant progress is likely if engineering methodologies \cite{DoE,sixsigma} can be leveraged to establish processes that reliably and repeatedly produce devices and subsystems with high-yield and deterministic performance. Key to such approaches is the ability to fabricate and characterize statistically significant numbers of devices, a major challenge when electrical measurements must be performed at milli-Kelvin temperatures in the presence of high magnetic fields.

Standard techniques for electrical test, such as the use of wafer-scale probe-stations, are challenging to implement in the deep-cryogenic environment. These difficulties are not fundamental barriers but rather technical, for instance, the challenge of connecting room temperature electronics to a large number of devices-under-test below \SI{1}{\kelvin}. A brute force approach, in which each device is independently connected to test electronics via its own wiring, becomes problematic for large wire-counts due to the thermal leak of the wiring itself \cite{Krinner2019}, the footprint of bulky connectors, and the likelihood of failure that stems from using meters of cabling across large temperature gradients. Together these aspects usually lead to device characterization proceeding via serial cool-downs of a dilution refrigerator, with each cycle taking several days.

Overcoming the challenge of multiple cool-downs, the need to perform high-throughput characterization has motivated previous work in realizing multiplexing devices and circuits. These include approaches that directly integrate the multiplexing switches into the quantum device chip \cite{Schaal2019,Eriksson,Smith} using the same technology. Embedding the multiplexer in this way however, increases the complexity and adds further steps to the fabrication process for quantum devices.

Here we describe a platform that decouples the quantum device and multiplexing circuit, fabricating a large-number of multiplexer chips via tape-out to a commercial CMOS foundry. This platform enables multiple quantum devices to be characterized in a single cool-down and with a standard cryostat wiring configuration. Our multiplexer (MUX) is based on custom cryo-CMOS technology and specifically designed for operation below \SI{100}{\milli\kelvin}. In the particular implementation reported here, 16 independent 1:5 MUX switches per die are configured by a room temperature microcontroller. The switches are based on parallel NMOS and PMOS transistors connected in a transmission gate (TG) topology, allowing for full rail-to-rail voltage swing of both the inputs and outputs. Our design further allows ad-hoc reconfiguration and daisy chaining of the multiplexer to suit particular device characterization needs. 

Demonstrating both control of high-impedance bias lines and low-impedance transport measurements, we tune a GaAs quantum dot to the few electron regime via the Cryo-CMOS MUX, and measure the Hall mobility across a 2'' InAs heterostructure. By making many parallel mobility measurements in a single cool-down we are able to map the variation in mobility across a wafer, enabling fast feedback between the growth of materials and device performance, an advantage for realizing and optimizing topological quantum devices.

\begin{figure*}
\includegraphics[width=\linewidth]{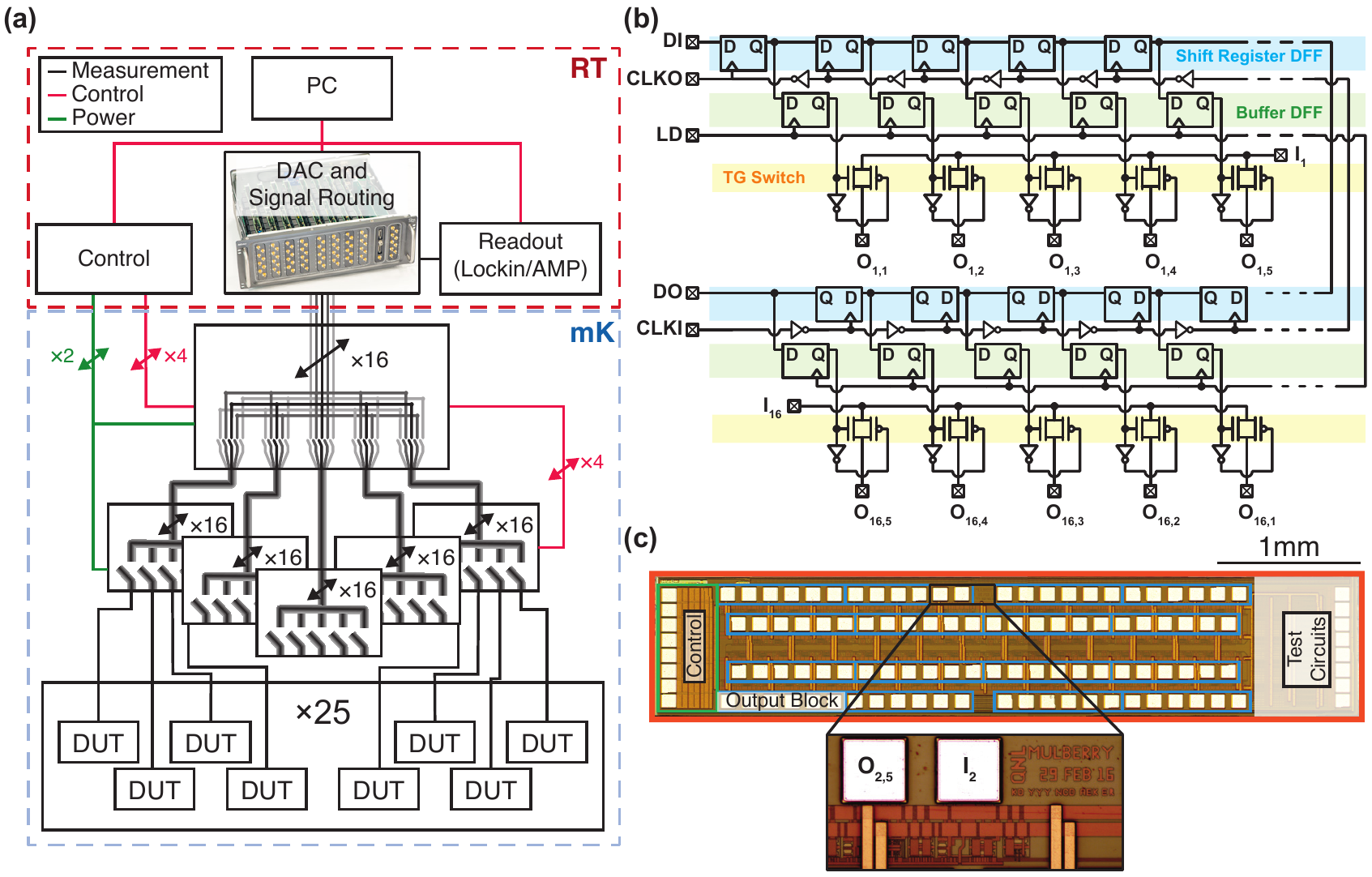}
\caption{\label{fig:fig1} (a) Experimental setup employing our Cryo-CMOS multiplexer chip to allow the simultaneous characterization of up to 25 DUTs at milli-kelvin temperature in a single cool-down. (b) Circuit block diagram of the multiplexer chip. Power supply pins $V_{DD}$ and $V_{SS}$ are not shown in the diagram, but can be set such that both negative and positive bias voltages may be applied, such that $V_{SS} <= V_{out} <= V_{DD}$. (c) Die photo of the fabricated chip. Inset shows part of the core switching circuit.}
\end{figure*}

\section{\label{sec:dd}Device Details}

A schematic of the experimental setup is shown in Fig.~\ref{fig:fig1} (a). Room temperature electronics are used to generate and measure voltages and currents in a circuit that comprises the device under test (DUT). The MUX chip is located at the milli-Kelvin stage of a dilution refrigerator, on a sample circuit board (PCB) adjacent to the DUT, and routes signals to and from it. Each MUX chip consists of 16 instances of a 1:5 analog multiplexing switch made from a \SI{0.35}{\micro\meter} AMS CMOS process \footnote{AMS SG. https://ams.com/. Shuttle run on 29 Feb, 2016}. The input of one multiplexer can be connected to any of the 5 outputs or disconnected entirely. MUX chips are controlled by a total of 6 control lines, including 2 power supply lines, and may be daisy chained together to increase the switch count. Software control over the switching topology is provided by a microprocessor \footnote{Cypress Semiconductor CY8C5888LTI-LP097} located at room temperature.

A circuit block diagram of the CMOS MUX is shown in Fig.~\ref{fig:fig1} (b). The multiplexing switches are of transmission gate (TG) topology, consisting of parallel NMOS and PMOS transistors of length \SI{0.7}{\micro\meter} and width \SI{40}{\micro\meter} and \SI{120}{\micro\meter} respectively. The TG switch structure allows for a maximum rail-to-rail input and output voltage swing of $V_{DD}-V_{SS} = \SI{3.3}{\volt}$. For our initial characterization, $V_{DD}$ is set to \SI{3.3}{\volt} and $V_{SS}$ is set to \SI{0}{\volt}, and are the MUX chip's high and low power supply respectively. We emphasize that by the use of lower values for $V_{SS}$ and $V_{DD}$, the MUX can be used to provide negative bias voltages. The use of a TG switch structure is necessary to ensure operation across the entire supply range. While using a single transistor switch design reduces the layout area and eliminates the need for complementary control signals, it limits the maximum output voltage swing to between $V_{SS}$ and $V_{DD}-V_{th,n}$ if a single NMOS transistor is used, or $V_{SS}+V_{th,p}$ and $V_{DD}$ if a single PMOS transistor is used, where $V_{th,n(p)}$ are the threshold voltage for NMOS (PMOS). The output swing can potentially become even more limited as the devices are cooled due to an increase in $V_{th,n(p)}$ as the intrinsic carrier density of the bulk, $n_{i}$, is exponentially dependent on temperature. As such, $n_i$ drops sharply at deep cryogenic temperatures ($< \SI{40}{\kelvin}$) due to bandgap widening and bulk carrier freeze out. This causes the bulk Fermi-level to increase, resulting in an increase in $V_{th,n(p)}$ \cite{BecEnz19,ghibaudo1997low}. The shift in threshold voltage, $\Delta V_{th,n(p)}$ depends on n-doping (p-doping) concentration. For the \SI{0.35}{\micro\meter} AMS process used for our devices, our measurements on transistor test structures indicate that $V_{th,n}$ increases from \SI{0.5}{\volt} to \SI{0.75}{\volt} as the chip is cooled from room temperature to $T = \SI{6}{\kelvin}$ (similar measurements were reported in \cite{DaoKass17}). The shift in $V_{th,p}$ is more severe, rising from \SI{0.8}{\volt} at \SI{300}{\kelvin} to \SI{1.35}{\volt} at \SI{6}{\kelvin}. Crucially, the threshold voltage of PMOS devices does not saturate at deep cryogenic temperatures but rather continues to increase as the temperature is lowered. 

An optical micrograph of the die, shown in Fig.~\ref{fig:fig1}(c), highlights the  key regions of the multiplexer. The power supply and digital control pins are placed on the left hand side of the chip, and the multiplexing circuit and IO pads are laid out in a symmetrical staggered arrangement taking up \SI[product-units = single]{3.1 x 0.8}{\milli\metre} of the core area. We note that the majority of this area is dominated by IO pads, whose area may be further miniaturized through the use of high-density interconnect technologies such as flip-chip bonding \cite{8429593}. The control interface is a simple shift register (edge triggered D flip-flop, DFF) bank with serial data (DI) and clock (CLKI) inputs from an external controller (see Fig.~\ref{fig:fig1} (b)). A second row of registers is used to buffer the switch controls. This enables the  switch configuration to be modified without affecting the state of the outputs. The change takes effect concurrently only at the rising edge of a global load (LD) signal. Notwithstanding comprehensive low temperature models and rigorous timing analysis, it was imperative to improve the circuit's immunity to timing violation by design. Therefore DI is shifted from left to right of the flip-flop chain whereas CLK is buffered inside each flip-flop and passed onto the cell to its left. Though the cost of this design is extra logic gates and delay, propagating the DI and CLKI signals from opposite direction guarantees proper logic operation because data always arrives before the rising edge of the clock. Further, the clocking events and subsequent logic changes in each flip-flop are then slightly staggered. This disperses the sudden draw of large load current from the power supply that would have otherwise occurred if all the flip-flops were clocked simultaneously. This is a pragmatic design, driven by the unique challenges of cryogenic measurements that involve power supply or digital signals being fed from room temperature instrumentation via meters of cables. In such setups, maintaining stringent delay matching or supply regulation can be difficult when compared to conventional room temperature bench-top configurations.   

By appropriate choice of topology, MUX chips may be chained together arbitrarily with no increase in the number of control lines. This is facilitated by the chaining of shift registers between chips, since the CLKI and DI pins are buffered and brought off chip on the clock out (CLKO) and data out (DO) pins respectively (see Fig.~\ref{fig:fig1} (b)). Power supply and LD signals may be shared between dies. For example, the chip may be converted to a $1:80$ multiplexer by tying all inputs together. Alternatively, by the use of 6 chips, it is possible to create $16 \times 1:25$ multiplexers in the arrangement shown in Fig.~\ref{fig:fig1} (a). Thus, the design is modular, allowing both the number of inputs and outputs to be increased without a corresponding increase in control lines, at the cost of a linear reduction at the rate that the state of the MUX may be changed.

\begin{figure}
\includegraphics[width=\linewidth]{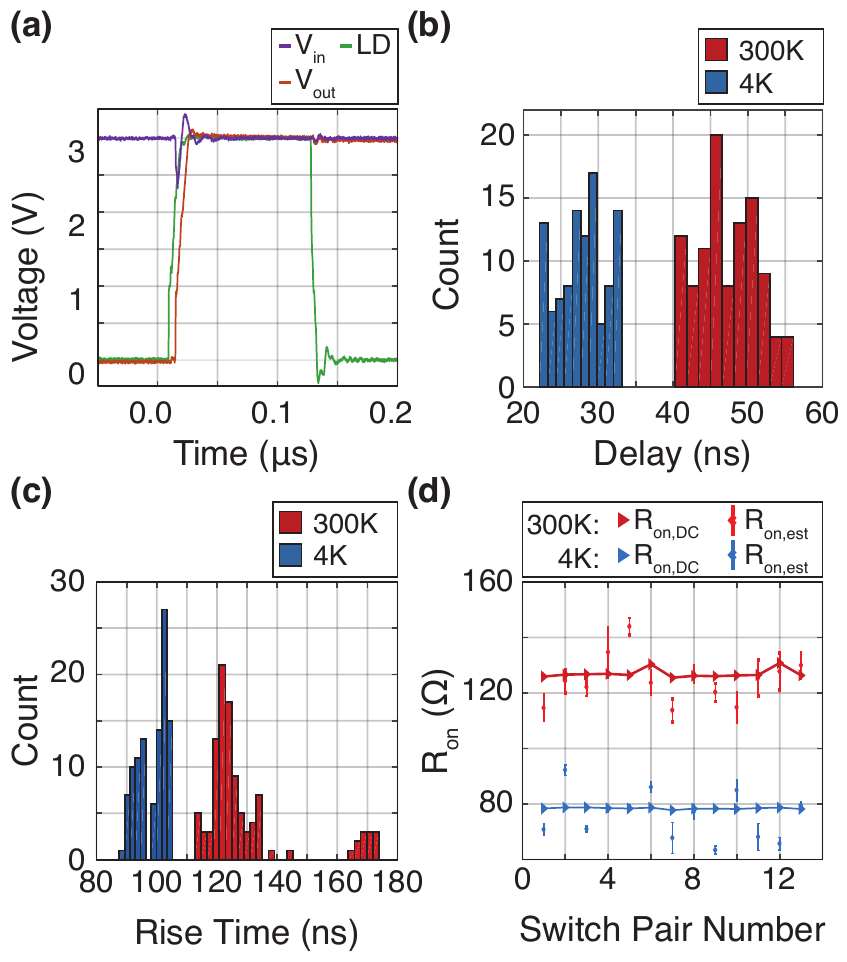}
\caption{\label{fig:fig2}  Characterization of the performance of the MUX chip at $T = \SI{300}{\kelvin}$ and \SI{4}{\kelvin}. (a) Multiplexing at \SI{4}{\kelvin} with a scope trace showing $V_{out}$ connecting to $V_{in} = \SI{3}{\volt}$ when the MUX switch is turned on. (b) Delay and (c) rise time through the MUX switches at $T = \SI{300}{\kelvin}$ and \SI{4}{\kelvin}. (d) On-state resistance estimated from delay and rise-time data, $R_{on,est}$, or measured directly using a lock-in amplifier, $R_{on,DC}$, at $T = \SI{300}{\kelvin}$ and \SI{4}{\kelvin}.}
\end{figure}

The switches in the MUX chip are designed to drive both high-impedence bias lines, such as the bias gates of a quantum dot device, as well as to probe transport phenomena through devices, while maintaining a low on-state resistance ($< \SI{200}{\ohm}$). The actual on-state resistance of the TG switch will be a function of applied source and drain voltage, $V_{SD}$, of the transistors. In steady state operation, for both biasing and transport applications, the transistors in the on-state are expected to have a negligibly small drain-source voltage, $V_{DS} \ll \SI{10}{\milli\volt}$, where they are designed operate in the ohmic region. The on-state resistance of the transistor can be expressed as,
\begin{equation}
    R_{on} \sim \frac{W}{\mu C_{ox} L (V_{GS}-V_{th})}
\end{equation}
where $\mu$ is carrier mobility, $C_{ox}$ is gate capacitance per unit area, $W$, $L$ are transistor width and length, and $V_{GS}$ is gate to source voltage. The effect of $V_{DS}$ is negligible in the limit that $V_{GS} - V_{th} \gg V_{DS}$.

At fixed temperature $R_{on}$ is a non-linear function of input voltage $V_{in}$, and supply voltages $V_{DD}$ and $V_{SS}$. Furthermore, the competing effect of increased $V_{th}$ and $\mu$ with cooling will cause $R_{on}$ to vary depending on operating temperature \cite{das2014effect,ghibaudo1997low}. The ratio of on-resistance at $T = \SI{300}{\kelvin}$ and \SI{4}{\kelvin}, $\gamma = R_{on,4K}/R_{on,300K}$, for both NMOS and PMOS transistors operating in the ohmic region can be expressed as: 
\begin{equation}\label{eq:rhon}
    \gamma_{n} = \frac{\mu_{n,300K}}{\mu_{n,4K}}\times\frac{1}{1 - \dfrac{\Delta V_{th,n}}{VDD-V_{in} - V_{th,n,300K} }}
\end{equation}
\begin{equation}\label{eq:rhop}
    \gamma_{p} = \frac{\mu_{p,300K}}{\mu_{p,4K}}\times\frac{1}{1 - \dfrac{\Delta V_{th,p}}{V_{in} - \lvert V_{th,p,300K}\rvert}}
\end{equation}
Here $\Delta V_{th,n(p)}$ is defined as $V_{th,n(p),4K} - V_{th,n(p),300K}$. The operating temperature and corresponding transistor type is denotes in the symbol subscript. We measure $\mu_{p,4K}$ to have increased by a factor of 2.4, whereas $\mu_{n,300K}$ is measured to be 4 times higher compared to its room temperature value \cite{DaoKass17}. Lower $\mu$ together with higher $V_{th}$ will cause $R_{on,p,4K}$ to be much higher than $R_{on,n,4K}$. 

To experimentally verify the performance of the Cryo-CMOS MUX chip at cryogenic temperatures, we wirebond the dies to a test PCB for measurement at $T = \SI{300}{\kelvin}$ and \SI{4}{\kelvin}. The outputs of 13 pairs of switches were wirebonded in series to measure the performance of the switches through the cryostat. The supply voltages for the test were set to $V_{DD} = \SI{3}{\volt}$ and $V_{SS} = \SI{0}{\volt}$ respectively. In addition, a pair of lines were shorted directly on the PCB, and used to calibrate the fixed resistance and delay through the cryostat. A representative trace, showing the output voltage $V_{out}$ being pulled high as the switch is closed, is shown in Fig.~\ref{fig:fig2} (a). A slight delay between the LD pin being asserted and $V_{out}$ being pulled high is visible, as well as a finite rise time which is set by the resistance of the switch and capacitance of wiring through the fridge. Histograms of delay and rise time across multiple switches in shown in Fig.~\ref{fig:fig2} (b) and Fig.~\ref{fig:fig2} (c) respectively at both \SI{300}{\kelvin} (red), and \SI{4}{\kelvin} (blue). A faster response is observed at \SI{4}{\kelvin}, which we attribute to the reduction of $R_{on}$, leading to a faster switching time. 

Finally, we measure the on-state resistance, $R_{on}$, of the switches in two ways. First, we directly measure the on-state resistance of the switches using conventional lock-in techniques, with a \SI{100}{\micro\volt} excitation, through a shorted pairs of switches. To find the individual switch resistance, $R_{on,DC}$, the previously calibrated resistance of the cryostat wiring is subtracted, and the result is divided by 2. Second, we extract the on-state resistance from the measurement of rise time across the 13 pairs, using the measured cable capacitance and known impedance of the measurement equipment, which we denote $R_{on,est}$. The measured values across all switches is plotted in Fig.~\ref{fig:fig2} (d), with both techniques yielding values in good agreement with each other.

\section{\label{sec:exp}Experimental Applications}
\subsection{Quantum Dots}
\begin{figure}
\includegraphics[width=\linewidth]{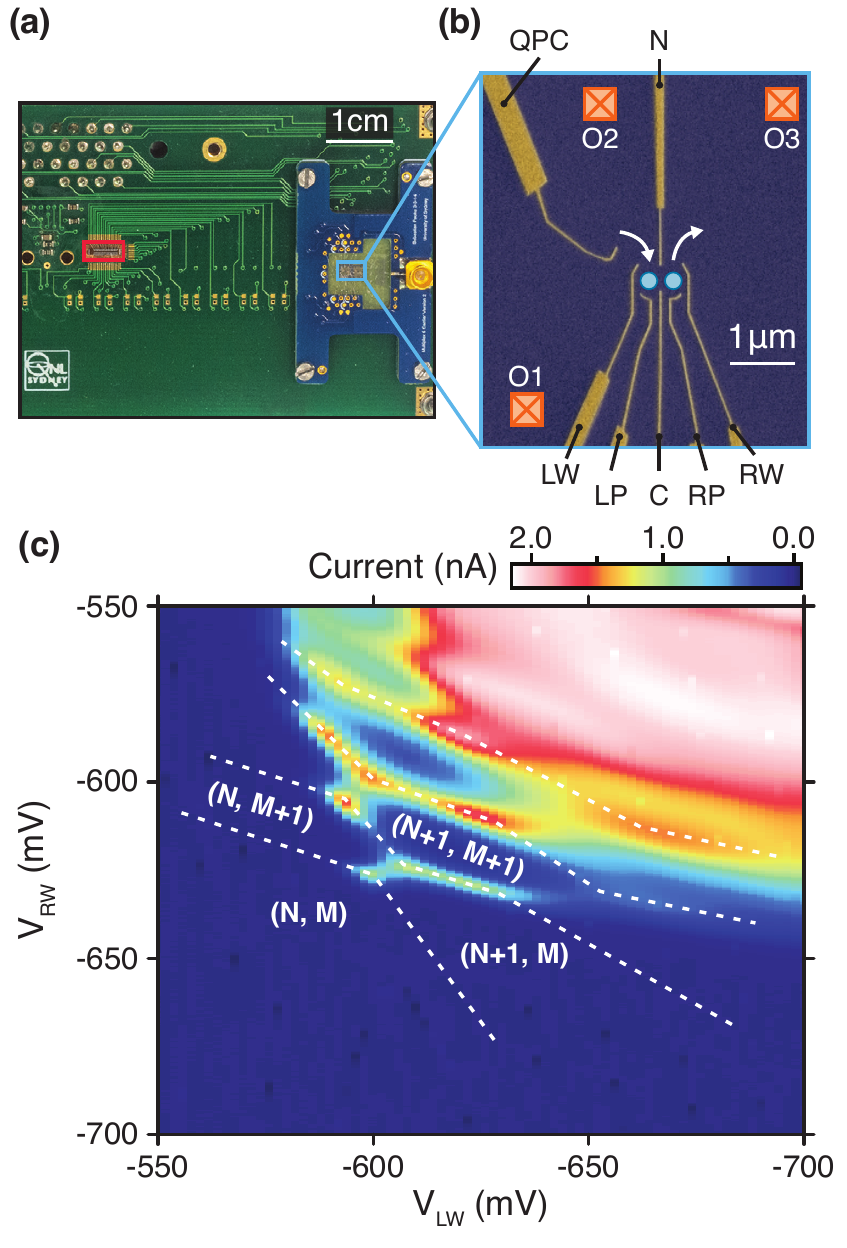}
\caption{\label{fig:fig3} (a) Photograph of the multiplexer characterization PCB showing the MUX chip (red), and an equivalent quantum dot connected to filtered DC lines via the MUX chip (blue). (b) False-color SEM of a quantum dot device, fabricated on a GaAs/(Al,Ga)As heterostructures. Ti/Au surface gates (gold) are used to define two quantum wells (blue dots). Locations of contacts to 2-dimensional electron gas are indicated in orange boxes. (c) Charge stability diagram of a double quantum dot showing the characteristic honeycomb pattern of a double quantum dot. White dashed lines indicate charge transitions and are a guide to the eye.}
\end{figure}
\begin{figure*}
\includegraphics[width=\linewidth]{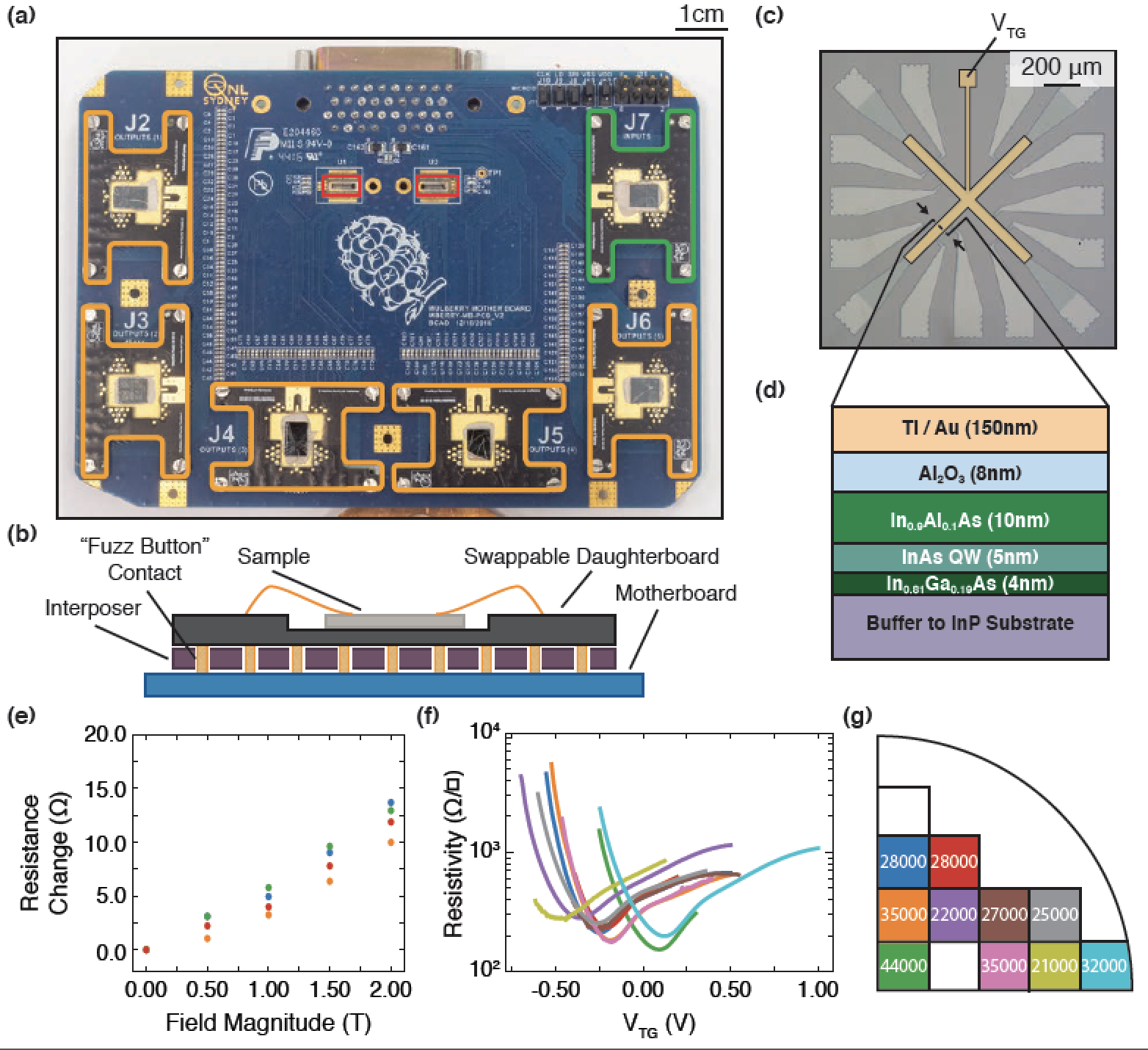}
\caption{\label{fig:fig4} (a) Photograph of multiplexed characterization PCB showing MUX die (red) with 5 daughter-boards (orange) allowing measurements, and an additional daughter-board (green), that has a connection to the fridge wiring bypassing the MUX chip. (b) Cross-section of mother-board, interposer and swappable daughter-board. (c) Optical micrograph of Hall bar device for characterizing mobility. A top gate (gold) allows for tuning of the density in the device. (d) Cross section of measured InAs heterostructure. (e) Resistance of shorted MUX lines as a function of perpendicular magnetic field, measured across four separate pairs of switches. (f) Hall resistivity as a function of $\textrm{V}_\textrm{TG}$ for n=9 Hall bar devices, measured across two cool-downs. (g) Extracted mobility as a function of position on growth wafer.}
\end{figure*}
In order to determine the suitability of our multiplexer for quantum device characterization, we connect it to the gates and ohmic contacts of a quantum dot device and perform transport measurements. A typical quantum dot device will have charging energies of around \SI{100}{\micro\electronvolt} \cite{PhysRevApplied.10.044058}. For a well defined quantum dot to be formed, the electrical noise and heat introduced by our proximal Cryo-CMOS multiplexer must be negligible relative to the charging energy \cite{RevModPhys.79.1217}, thus quantum dot measurements provide a means of determining the suitability of our MUX for purpose. The quantum dot device is fabricated on an MBE grown GaAs/(Al,Ga)As heterostructure, which forms a 2-dimensional electron gas (2DEG) \SI{91}{\nano\meter} below the surface. TiAu gates are patterned on the surface to define the dots, separated by an \SI{10}{\nano\meter} \ce{HfO2} dielectric. An optical micrograph of the device is shown in Fig.~\ref{fig:fig3} (a), with the MUX chip highlighted in the red box and the quantum dot device highlighted in the blue box. A false-color SEM of a similar device is shown in Fig~\ref{fig:fig3} (b). The sample is mounted at the milli-Kelvin stage of a dilution refrigerator with a base temperature of \SI{8}{\milli\kelvin}. Negative voltages are applied to the surface gates (gold) to create quantum dots (blue) containing a discrete number of electrons. The occupancy of each dot is denoted $(N, M)$ where $N$ ($M$) is the number of electrons on the left (right) quantum dot. Current is passed through the quantum dot via contacts to the 2DEG, O2 and O3, where current is only able to flow when there are available electron states in both the left and the right quantum dot. In order to allow negative bias voltages to be used, the MUX chip is operated with $V_{SS} = \SI{-2}{\volt}$ and $V_{DD} = \SI{1}{\volt}$.

In Figure 3(c), the current through the quantum dot is shown, with both gates and ohmic contacts routed through the Cryo-CMOS MUX chip. Conventional lock-in techniques are used for current readout with an excitation voltage of \SI{100}{\micro\volt} across the device. A honeycomb pattern characteristic of a double quantum dot is visible, with charge transitions indicated by dotted lines. Similar measurements were carried out for different gate configurations. In the steady-state we find the additional heat or noise generated by the MUX chip to be negligible, with the base temperature of the cryostat unaffected by the multiplexer.

\subsection{Mobility Characterization}
Finally, we demonstrate the use of our multiplexer for performing batch materials characterization of InAs heterostructures, of interest for the purpose of realizing topological qubits \cite{s41578-018-0003-1}. Here we are focused on determining how parameters such as the carrier mobility of the electron gas varies across a wafer, a key metric for device performance \cite{manfra_hmob}. To carryout batch-style measurements we have developed a device packaging approach that allows many quantum devices to be bonded onto separate daughter PCBs \cite{CollessRSI}, that are collectively mounted on a motherboard that also houses the CMOS MUX chips, as shown in Fig. 4(a). This setup allows mounting of five dies, each containing 4 devices, in the dilution refrigerator in a single cool-down, with an additional sixth die configured such that its electrical connections bypass the cyro-CMOS MUX. The daughter-boards are connected to the motherboard via an interposer, that allows for samples to be rapidly interchanged in between  cool-downs \cite{CollessRSI} (Fig.~\ref{fig:fig4} (b)). 

In our demonstration the heterostructure consists of a InAlAs/InAs/InGaAs quantum well grown \SI{10}{\nano\meter} below the surface on a 2" (100) InP substrate (Fig.~\ref{fig:fig4} (c-d)). An \SI{8}{\nano\meter} layer of Al was deposited in-situ to induce superconductivity in the quantum well via the proximity effect. Hall bar devices were defined across a quarter wafer of heterostructure, and the Al removed in the active area both by standard wet etching techniques. A global ALD \ce{Al2O3} gate dielectric was then deposited, followed by a Ti/Au top gate defined by e-beam lithography. Magnetoconductance measurements of the Hall bar were performed simultaneously as a function of top gate voltage in a perpendicular magnetic field using standard lock-in techniques. Fig.~\ref{fig:fig4} (f) shows the extracted resistivity as a function of top gate voltage for 9 samples, obtained across two cool-downs. No degradation in device performance is seen compared to devices measured without the Cryo-CMOS MUX in-line. Using this technique, we can map out the mobility as a function of position of the die on the growth wafer, as indicated in Fig.~\ref{fig:fig4} (g). We observe that dies coming from the edge of the wafer suffer a degradation of mobility by a factor of two (\mob{27000}) compared to dies near the center of the wafer (\mob{44000}).The magnetoconductance measurements also allow us to study the additional inline resistance that emerges as a function of magnetic field. As shown in Fig.~\ref{fig:fig4} (e), we observe an additional linear resistance of \SI{6}{\ohm\per\tesla} within the magnetic field range of \SIrange{-2}{2}{\tesla} studied in this work.\\

\textit{Note:} In the final stage of submitting this manuscript a preprint describing similar work using commercial off-the-shelf CMOS multiplexing circuits has appeared \cite{Wuetz}.

\begin{acknowledgments}
This research was supported by Microsoft Corporation and the ARC Centre of Excellence for Engineered Quantum Systems. We thank Y. Yang, X. Croot, and A. Moini for technical assistance and useful discussions. We acknowledge the facilities as well as the scientific and technical assistance of the Research \& Prototype Foundry Core Research Facility at the University of Sydney, part of the Australian National Fabrication Facility (ANFF), and the NSW node of ANFF at the University of New South Wales.
\end{acknowledgments}

\bibliography{mulberry}
\end{document}